\documentclass{article}

\usepackage{amsmath,amssymb}
\usepackage{latexsym}

\begin{document}

\newtheorem{thm}{Theorem}[section]
\newtheorem{lem}[thm]{Lemma}
\newtheorem{col}[thm]{Corollary}
\newtheorem{df}[thm]{Definition}
\newtheorem{exam}[thm]{Example}
\newtheorem{rem}[thm]{Remark}
\newtheorem{Rem}{Remark.}
\renewcommand{\theRem}{}
\newtheorem{claim}[thm]{Claim}
\newtheorem{Claim}{Claim.}
\renewcommand{\theClaim}{}
\newenvironment{proof}{\noindent{\bf Proof.}}{\hfill $ \square $}

\renewcommand{\theequation}{\fnsymbol{equation}}

\newcommand{\z}{\mathrm Z}
\newcommand{\f}{\mathbf f}
\newcommand{\Max}{\mathrm{max}_2}

\title{ A new function algebra of EXPTIME functions by safe nested
recursion }

\author{ Toshiyasu Arai and Naohi Eguchi \\ \\
         Graduate School of Engineering \\
         Kobe University}

\date{}

\maketitle

\begin{abstract}
Bellantoni and Cook have given a function-algebra characterization of the
 polynomial-time computable functions via an unbounded recursion scheme
 which is called safe
 recursion. 
Inspired by their work, we characterize the exponential-time computable
 functions with the use of
 a safe variant of nested recursion.
\end{abstract}

\section*{Introduction}

Function algebras for complexity classes have been investigated
with an interest in what kind of recursion scheme captures which function
complexity class.
Nowadays we know at least two different ways to this problem.
On the one hand, there is the approach by \emph{bounded} recursion.
A memorable contribution to a computational complexity class was given
by A. Cobham.
In his 1965 paper \cite{cob}, he characterized the class of the polytime
functions, using  
bounded recursion on notation.
It is known that the same operation of bounded recursion generates various
complexity classes with the presence of 
special initial functions like the smash function
$ \# ( x , y ) = 2^{ | x | \cdot | y | } $.
For example, see Clote \cite{clote99}.

On the other hand, 
there is the approach using \emph{safe} recursion, 
which requires no explicit bounding.
In 1992, S. Bellantoni and S. Cook introduced a class 
$\mathcal B $ which is
closed under the scheme of safe recursion on notation.
In functions of $ \mathcal B $,
variables are distinguished as to their positions.
Variables $\vec x$ occurring to the left of the semi-colon are called
\textit{normal}, 
and variables $ \vec a $ to the right are called \textit{safe}:
\[
  f( \ \underbrace{ x_{1} , \dots ,x_{k} }_{\mbox{normal } } \ ; \ \underbrace{
      a_{1} , \dots ,a_{l} }_{ \mbox{safe} } \ ) \in \mathcal B^{k,l} 
\]
Roughly speaking, 
the normal positions are used only for recursion,
while the safe positions are used only for substitution.
Let $\mathcal F_{\mathrm P}$ be the class of functions computed by a
deterministic 
Turing machine in polynomial time of the binary lengths of inputs, 
and $\mathcal B_{normal}$ be 
$ \bigcup_{ k \in \omega } \mathcal B^{k,0}$.
In \cite{bel_cook} it has been shown that

\[
  f( \vec{ x } ; ) \in \mathcal B_{normal} \Longleftrightarrow f(
      \vec{x}) \in \mathcal F_{\mathrm P} .
\]
The class $ \mathcal B $ contains as initial functions only specific elementary 
ones with low growth-rates.
Hence we consider that the safe recursion scheme grasps the class 
$ \mathcal F_{ \mathrm P } $ well.

Since the Bellantoni-Cook characterization, 
many function complexity classes have been characterized in similar manners by
some safe representations,
e.g., LINSPACE, LOGSPACE, NC, levels of the polynomial hierarchy or ETIME, 
cf. Bellantoni \cite{bel92} or Clote \cite{clote97, clote99}.
In the spirit of Bellantoni and Cook \cite{bel_cook}, 
we characterize the class of EXPTIME functions.
Let $ \mathcal F_{ \mathrm{EXP}}$ be the class of functions computed by
a deterministic Turing machine in exponential time.
In this paper, we introduce a class 
$ \mathcal N $ in the Bellantoni-Cook style such that
\[
  f( \vec{ x } ; ) \in \mathcal N_{normal} \Longleftrightarrow f(
      \vec{x}) \in \mathcal F_{\mathrm{EXP}}.
\]

The class $\mathcal B$ is generated from the following initial functions 
and operations:
\begin{description}
\item[Zero] $O ( \vec x ; \vec a ) = 0$
\item[Projections]
$I^{ k, l }_j ( x_1, \dots , x_k ; a_1, \dots a_l ) = 
 \left\{
 \begin{array}{ll}
    x_j 
  & \text{if $ 1 \leq j \leq k $,} \\
    a_{ j - k }
  & \text{if $ k < j \leq k + l $.}
 \end{array}
 \right.$
\item[Successors] $S_i (;a) = 2 a + i \quad ( i = 0,1)$
\item[Predecessor] $P (;a) = \lfloor a / 2 \rfloor$
\item[Conditional]
$C ( ; a , b , c ) =
 \left\{
  \begin{array}{ll}
    b
  & \text{if $ a \equiv 0 \mod 2 $,} \\
    c 
  & \text{else.}
  \end{array}
 \right.$
    \item[Safe composition] If
$ h_1 , \dots , h_m \in \mathcal B^{ k , l } $,
$ g \in \mathcal B^{ n , m } $ and 
$ i_1 , \dots , i_n \in \{ 1, \dots , k \} $, 
then $ f \in \mathcal B^{ k, l } $ is defined by
\begin{equation}
 f ( \vec x ; \vec a )
 = g ( x_{ i_1 } , \dots , x_{ i_n } ; h_1 ( \vec x ; \vec a ) , \dots ,
                                       h_m ( \vec x ; \vec a )
     ) .
 \label{scomp}
\end{equation}
    \item[Safe recursion on notation (SRN)] If 
$g \in \mathcal B^{ k , l }$ and $ h_0 , h_1 \in \mathcal B^{ 1 + k , l + 1 } $,
then $ f \in \mathcal B^{ 1 + k , l } $ is defined by 
\begin{equation} 
\left\{
\begin{array}{rcl}
f( 0, \vec x ; \vec a ) & = & g( \vec x ; \vec a ), \\
    f(S_0 ( ; y ), \vec x ; \vec a) 
         & = & h_0 (y, \vec x ; \vec a , f(y , \vec x ; \vec a)) , 
\ \text{provided $ y \neq 0 $,} \\ 
    f ( S_1 ( ; y ) , \vec x ; \vec a ) & = &
    h_1 ( y , \vec x ; \vec a , f ( y , \vec x ; \vec a ) ) .
\end{array}
\right.
\label{srn}
\end{equation}
\end{description} 
\ \\
\textbf{Notations.}
$ S_i ( ; a ) $ is denoted by $ a i $, 
since $ S_i ( ; a ) \equiv ( a i )_2 $
if $ a $ is written in the binary representation.
Let $ \oplus $ denote the \emph{concatenation} as
$ x \oplus y 
  = ( x_n \cdots x_1 y_m \cdots y_1 )_2
$
for $ x = ( x_n \cdots x_1 )_2 $ and $ y = ( y_m \cdots y_1 )_2 $,
$ |x| $ the length of the binary representation of $x$, 
i.e., 
$ |x| = \lceil \log_2 (x+1) \rceil $,
which is called the binary length of $x$
or the length of $x$ in short.
And, for 
$ \vec x = ( x_1 , \dots , x_n ) $, let
$ | \vec x | := ( |x_1| , \dots , |x_n| )$ and
$ \max \vec x := \max \{ x_i : i=1, \dots , n  \} $.

\begin{Rem}
\label{rem} \normalfont
The safe composition scheme (\ref{scomp}) is introduced 
in Handley and Wainer \cite{handley}.
This scheme is a restriction of Bellantoni and
Cook's one in \cite{bel_cook}:
\begin{quotation}
If
$t_1, \dots ,t_n \in \mathcal B_{normal} , h_1, \dots ,h_m \in \mathcal
	   B $ and $ g \in \mathcal B $, 
then $f \in \mathcal B $ is defined by 
$ f( \vec x ; \vec a )=g( t_1 ( \vec x ;) , \dots , t_n ( \vec x ;) 
       ; h_1 ( \vec x ; \vec
	   a ) , \dots , h_m ( \vec x ; \vec a )).
$  
\end{quotation}
Handley and Wainer have proved that the
same class is obtained by the restricted scheme
over unary notation.

The length of a function in $ \mathcal F_{ \mathrm P } $ is
bounded by some polynomial on the lengths of inputs.
Polynomials are closed under composition,
so is $ \mathcal F_{ \mathrm P } $. 
This is consistent with the Cobham characterization of 
$ \mathcal F_{ \mathrm P } $ in \cite{cob}.

As well known, 
the length of a function $f (\vec x)$ in $ \mathcal F_{ \mathrm{ EXP } } $
is bounded by 
$ 2^{ p ( | \vec x | ) } $ for some polynomial $ p ( \vec x ) $.
Nevertheless
$ 2^{ p ( \vec x )
    }
$
functions are not closed under composition
as seen from a simple example such that
$ f ( f ( x)) = 2^{ 2^x } $ for $ f ( x ) = 2^x $.
Therefore the class $ \mathcal F_{ \mathrm{ EXP } } $
is \emph{not} closed under composition either.
Hence we need to restrict the safe composition scheme in \cite{bel_cook}.
However, it turns out that substituting
$ \vec t ( \vec x ; ) \in \mathcal B $
into normal arguments yields no new functions.
Namely,
if
$t_1, \dots ,t_k \in \mathcal B_{normal} , h_1, \dots ,h_l \in \mathcal
	   N $ and $ g \in \mathcal N $, 
then  
\[
 f( \vec x ; \vec a )=g( t_1 ( \vec x ;) , \dots , t_k ( \vec x ;) 
       ; h_1 ( \vec x ; \vec
	   a ) , \dots , h_l( \vec x ; \vec a ))
\]
belongs to $ \mathcal N $.
Similar considerations are seen in Ritchie \cite{ritchie}.
\end{Rem}

\section{Safe nested recursion and a class $ \mathcal N $ }
\begin{df} \label{def1}
\normalfont
$\mathcal N$ is the smallest class containing the initial functions of 
$\mathcal B$ 
and closed under safe composition and safe nested recursion on notation
(SNRN),
which is defined in Definition \ref{def3} below.
\end{df}

Computations of functions defined by nested recursion run along 
the lexicographic ordering.
(For the general definition of nested recursion, see Rose \cite{rose}.)
We weaken it and define
$ ( v_1, \dots , v_k ) \prec ( y_1, \dots , y_k ) $
for any $ k \geq 1 $.

\begin{df}[$\prec$-predecessors] 
\label{def2} \normalfont \
If
$ 1 \leq n \leq k $, 
$y_n \neq 0 $, and
$ v_{ n+1 } , \dots , v_k \in
 \{ y_i : 1 \leq i \leq k
 \} \cup \{ P ( ; y_i ) : 1 \leq i \leq k
         \}
$, then
\[
 ( y_1 , \dots , y_{ n-1 } , P(;y_n), v_{n+1}, \dots , v_k
 ) \prec
 ( y_1 , \dots , y_{ n-1 } , y_n, y_{n+1}, \dots , y_k ) .
\]
If $ \vec v \prec \vec y $, 
then we call $ \vec v $ a $\prec$-\emph{predecessor} of $ \vec y $.
\end{df}

Given $ \vec y$, 
a $ \prec$-predecessor of $ \vec y$ is not, in general, uniquely determined.
Thus we introduce the \emph{$\prec$-functions}.
A $\prec$-function $ \f$ indicates which predecessor should be chosen.
The choice of a predecessor, however, does not depend on the value
itself of $ \vec y$, but on the configuration or \emph{type} of $ \vec y$.
Hence we define the type $ \tau ( \vec y)$ of $ \vec y$.
\begin{df}[Types, $\prec$-functions]
\normalfont \
\begin{enumerate}
\item (Types)
We fix the signature $\Sigma=\{0, 1, \z \}$.
Let $\Sigma^k$ be the set of words of length $k$ consisting of elements
      of $\Sigma$.
Then the \emph{type} $ \tau (y_1, \dots, y_k)$ of 
$(y_1, \dots, y_k)$ 
is inductively defined by
  \begin{itemize} \def\labelitemi{--}
  \item $ \tau (y0) = 0 \ (y \neq 0)$, $ \tau (y1) = 1$, $ \tau (0) =
	 \mathrm Z $, and
  \item $ \tau ( \vec y) = \tau (y_1, \dots, y_k) = 
          \tau (y_1) \cdots \tau (y_k)$.
  \end{itemize}
And we set $\Sigma^k_0 := \Sigma^k \setminus \{ \z \cdots \z
      \}$.
By the definition of $\tau$, 
\[
 \tau ( \vec y) \in \Sigma^k_0 \Longleftrightarrow \max \vec y \neq 0.
\]
Therefore, 
$ \Sigma^k_0 = \{ \tau ( \vec y) : \max \vec y \neq 0 \}$
for $ \vec y = (y_1, \dots, y_k)$.
\item ($\prec$-functions)
To define $\prec$-functions,
we introduce the \emph{modified projection} functions $J^k_j$ 
$(1 \leq j \leq 2k)$ defined by
\begin{equation*}
  J^k_j (x_1, \dots, x_k) =
  \begin{cases}
  x_j & \text{if $1 \leq j \leq k$,} \\
  \lfloor x_{j-k} / 2 \rfloor & \text{if $k < j \leq 2k$.}
  \end{cases}
\end{equation*}
Then a function
$ \mathbf f : \{1, \dots, k\} \times \Sigma^k_0 \rightarrow 
              \{1, \dots, 2k\}
$
is called a \emph{$\prec^k$-function} iff
for all $ \vec y = (y_1, \dots, y_k) \neq (0, \dots, 0)$,
\[
  J^k_{ \f( \tau ( \vec y))} ( \vec y) :=
  (J^k_{ \mathbf f (1, \tau ( \vec y))} ( \vec y), \dots,
   J^k_{ \mathbf f (k, \tau ( \vec y))} ( \vec y)
  ) \prec \vec y.
\]\end{enumerate}
\end{df}

\begin{exam} \label{exam_pred} \normalfont
Let us consider the cases $k=1, 3$ for $ \vec y = (y_1, \dots, y_k)$. 

\textsc{Case} $k=1$.
The only $\prec$-predecessor of $x0$ and $x1$ is $x$.
Hence the only possible choice of the $\prec^1$-function $ \mathbf f$ is
$ \mathbf f (1, \sigma ) = 2$ for each 
$ \sigma \in \{0, 1\} = \Sigma^1_0$. 

\textsc{Case} $k=3$.
Consider the following function 
$ \f : \{1, 2, 3\} \times \Sigma^3_0 \rightarrow 
       \{1, \dots, 6 \}
$:
\begin{equation*}
\left\{
  \begin{array}{lllllll}
  (1, \sigma_1 \sigma_2 i) &\mapsto& 1, \quad 
  (2, \sigma_1 \sigma_2 i) &\mapsto& 2 , \quad
  (3, \sigma_1 \sigma_2 i) &\mapsto& 6 , \\
  (1, \sigma i \z) &\mapsto& 1         , \quad
  (2, \sigma i \z) &\mapsto& 5         , \quad
  (3, \sigma i \z) &\mapsto& 1         , \\
  (1, i \z \z) &\mapsto& 4  , \quad
  (2, i \z \z) &\mapsto& 2  , \quad
  (3, i \z \z) &\mapsto& 3  
  \end{array}
\right.
\end{equation*}
for each $i=0, 1$ and each $ \sigma, \sigma_1, \sigma_2 \in \Sigma $.
Then the following $\prec$-predecessors in the LHS are 
$J^3_{ \f ( \tau ( \vec y))} ( \vec y)$ of the RHS 
$ \vec y = (y_1, y_2, y_3)$: 
\begin{equation*}
\left\{
  \begin{array}{lll}
  (x, y, z) & \prec & (x, y, zi), \\
  (x, y, x) & \prec & (x, yi, 0), \\
  (x, 0, 0) & \prec & (xi, 0, 0)
  \end{array}
\right.
  (i=0, 1)
\end{equation*}
Therefore, $ \f$ is a $ \prec^3$-function.
\end{exam}

Now we define safe nested recursion on notation.
In Definition \ref{def3}, let 
$ f ( \vec x ; \vec a, \vec b ) [ \vec g ( \vec y ; \vec c ) / \vec b] $
denote 
$ f ( \vec x ; \vec a, \vec g ( \vec y ; \vec c))$,
the result of simultaneous substitution.
And, for functions $f_{w, 1}, \dots, f_{w, l}$, 
let $ \vec f_w ( \vec x; \vec a)$ abbreviate 
$(f_{w, 1} ( \vec x; \vec a), \dots, f_{w, l} ( \vec x; \vec a))$.

\begin{df}[Safe nested recursion on notation (SNRN)]
\label{def3} \normalfont
Suppose that 
$ g \in \mathcal N^{m,l} $ and
$ h_w$,
$ t_{w, 1}, \dots, t_{w, l} $,
$ s_{w, 1}, \dots, s_{w, l} \in \mathcal N^{k+m,l+1} $
for \emph{each} $w \in \Sigma^k_0$.
Also suppose that 
$ \mathbf f_1$,
$ \mathbf f_2$ and
$ \mathbf f_3$ are $\prec^k$-functions.

Then $f \in \mathcal N^{k+m,l}$ is defined by 
\begin{equation}
\left\{
\begin{array}{rcll}
f ( \vec 0, \vec x; \vec a) &=& g ( \vec x; \vec a), & \\
f ( \vec y, \vec x; \vec a) &=&
  h_{ \tau ( \vec y)} ( \vec v_1, \vec x; \vec a, 
                         f ( \vec v_1, \vec x; \vec b )) &
\\
&&
[\vec t_{ \tau ( \vec y)} ( \vec v_2, \vec x; \vec a, 
   f ( \vec v_2, \vec x; \vec c)) / \vec b] & \\
&&
[\vec s_{ \tau ( \vec y)} ( \vec v_3, \vec x; \vec a,
   f ( \vec v_3, \vec x; \vec a)) / \vec c],  &
\text{provided $ \max \vec y \neq 0$,} 
\end{array}
\right.
\label{snrn}
\end{equation}
where, for every $j=1, 2$ and $3$, $ \vec v_j$ abbreviates
$ J^k_{ \mathbf f_j ( \tau ( \vec y))} ( \vec y)$,
and hence $ \vec v_j \prec \vec y$. 
\end{df}

\begin{exam}
\label{exam} \normalfont 
First consider the case $k=1$.
$\Sigma^1_0 = \{0, 1\}$ and the scheme (\ref{snrn}) runs as follows. 
\begin{equation*}
\left\{
\begin{array}{rcll}
f (0, \vec x; \vec a) &=& g ( \vec x; \vec a), & \\
f (y0, \vec x; \vec a) &=&
  h_0 (y, \vec x; \vec a, 
                         f (y, \vec x; \vec b )) &
\\
&&
[\vec t_0 (y, \vec x; \vec a, 
   f (y, \vec x; \vec c)) / \vec b] & \\
&&
[\vec s_0 (y, \vec x; \vec a,
   f (y, \vec x; \vec a)) / \vec c], & 
\text{provided $y \neq 0$,} \\
f (y1, \vec x; \vec a) &=&
  h_1(y, \vec x; \vec a, 
                         f (y, \vec x; \vec b )) & \\
&&
[\vec t_1 (y, \vec x; \vec a, 
   f (y, \vec x; \vec c)) / \vec b] & \\
&&
[\vec s_1 (y, \vec x; \vec a,
   f (y, \vec x; \vec a)) / \vec c] &
\end{array}
\right.
\end{equation*}
Taking the projection function $I^{1+m,l+1}_{1+m+i}$
as both $t_{0, i}$ and $t_{1,i}$,
this scheme is identical to the scheme (\ref{srn}) of SRN, 
yielding $ \mathcal B \subset \mathcal N $.

Using the SNRN operation, we can define complex exponential functions
 step by step.
The following construction is crucial in Section \ref{sec2}.

Let $g (;a) := S_0 (;a) = a0$.
\begin{enumerate}
\item
\label{exam_0}
$ 
 f_0 ( x ; a ) = 
    2^{
        2^{
            | x |
          }
      }
 \cdot a.
$

For each $i \in \{0, 1\} = \Sigma^1_0$,
take $I^{1, 2}_3$ as $h_i$, $I^{1, 2}_3$ as $t_i$ and $I^{1, 2}_2$ as
     $s_i$.
Then the following equations define $f_0$.
\begin{eqnarray*}
   f_0 ( 0 ; a ) & = & 2 a = a 0 , \\
   f_0 ( x i ; a ) & = & 
     2^{ 
         2^{
             |x| + 1
           }
       }
     \cdot a
   =
     2^{
         2^{
             |x|
           }
       }
     \cdot
     (
       2^{
           2^{
               |x|
             }
         }
       \cdot a
     ) \\
   & = & 
      f_0 ( x ; f_0 ( x ; a )) . \qquad (i=0, 1)
\end{eqnarray*}
\item
\label{exam_1}
$ f_1  (x, y, z; a) = 
   2^{
       2^{
           |x| \cdot |y| + |z|
         }
     }
   \cdot a.
$

Take $I^{3, 2}_5$ as $h_w$ for every $w \in \Sigma^3_0$.
Take $I^{3, 2}_4$ as $t_{ \sigma i \z}$ and $t_{i \z \z}$ for every 
$ \sigma \in \Sigma $ and $i \in \{0, 1\}$, and else $I^{3, 2}_5$ as $t_w$.
Take $I^{3, 2}_4$ as $s_w$.
And take the $\prec^3$-function $ \f$ in Example \ref{exam_pred}.
Then the following equations define $f_1$.
\begin{eqnarray*}
    f_1 (0, 0, 0; a)
&=& a 0, \\
    f_1 (x, y, z i; a)
&=& 2^{2^{|x| \cdot |y| + |z|+1}} \cdot a \\
&=& f_1 (x, y, z; f_1 (x, y, z; a)), \\
    f_1 (x, y i, 0; a)
&=& 2^{2^{|x| (|y|+1)}} \cdot a =
    2^{2^{|x| \cdot |y| + |x|}} \cdot a \\
&=& f_1 (x, y, x; a), \\
    f_1 (xi, 0, 0; a) 
&=& f_1 (x, 0, 0; a). \qquad (i=0, 1)
\end{eqnarray*}
\item
\label{exam_2}
$f_2 (x, y, z, u, v, w; a) = 
 2^{2^{|x| \cdot |y| \cdot |z| +|u| \cdot |v| +|w|}} \cdot a.
$

As the former two cases, we define $f_2$ by
\begin{eqnarray*}
    f_2 (0, 0, 0, 0, 0, 0; a)
&=& a 0, \\
    f_2 (x, y, z, u, v, w i; a)
&=& f_2 (x, y, z, u, v, w; f_2 (x, y, z, u, v, w ; a)), \\
    f_2 (x, y, z, u, v i, 0; a)
&=& f_2 (x, y, z, u, v, u; a), \\
    f_2 (x, y, z i, u, 0, 0; a) 
&=& f_2 (x, y, z, x, y, 0; a), \\
    f_2 (x, yi, 0, u, 0, 0; a) 
&=& f_2 (x, y, 0, u, 0, 0; a), \\
    f_2 (xi, 0, 0, u, 0, 0; a)
&=& f_2 (x, 0, 0, u, 0, 0; a), \\
    f_2 (0, 0, 0, ui, 0, 0; a)
&=& f_2 (0, 0, 0, u, 0, 0; a). \qquad (i=0, 1)
\end{eqnarray*}
\end{enumerate}
As above, we can define more complicated exponential functions.
Therefore, suitable applications of safe composition yield
$2^{2^{p (| \vec x|)}} \in \mathcal N_{normal}$
for any polynomial $p ( \vec x)$. 
\end{exam}

\begin{rem} \label{rem2} \normalfont
Let us consider the scheme (\ref{snrn}) of SNRN.
It turns out that only
$\vec s_w ( \vec y, \vec x; \vec a, b)= 
 \vec s_w ( \vec y, \vec x; \vec a )
$
suffices to prove 
$ \mathcal F_{ \mathrm{EXP} } \subseteq \mathcal N_{normal} $.
The definition of SNRN suggests that 3-times nesting is allowed. 
Even if we admit any constant number of nestings,
the same class will be generated.
Nevertheless, 
we need only the above restricted scheme in later discussions. 
In addition, as seen in Section \ref{lastsec},
the same class is obtained even by replacement of $ \vec x $.
\end{rem}

\renewcommand{\theequation}{\arabic{equation}}
\setcounter{equation}{0}
 
\section{Simultaneous safe nested recursion}
\label{sec2}

In this section, 
we prove that the class $ \mathcal N_{normal}$ is closed under 
a scheme of simultaneous SNRN (Theorem \ref{thm2}).
Furthermore, we show that a substitution of a large value for a recursion parameter
is admitted (Corollary \ref{col}).

\begin{thm} \label{thm2}

Suppose that
$ f_1, \dots, f_l $
are defined from
$ h_1, \dots, h_l \in \mathcal N^{m,l}$ and
$\prec^k$-functions  
$ \f_1$ and
$ \mathbf f_2$
simultaneously by safe nested recursion on notation such that
for each $ j = 1, \dots , l $ and for 
$ \vec v_i = J^k_{ \f_i ( \tau ( \vec y))} ( \vec y)$ $(i=1, 2)$,
\begin{equation*}
 \begin{cases}
 f_j ( \vec 0 , \vec x ; \vec a ) = 
 h_j ( \vec x ; \vec a ),
 & \text{} \\
 f_j ( \vec y, \vec x ; \vec a ) =
 f_j ( \vec v_1, \vec x; 
       f_1 ( \vec v_2, \vec x; \vec a), \dots, 
       f_l ( \vec v_2, \vec x; \vec a)
     )
 & \text{$( \max \vec y \neq 0)$.} 
 \end{cases}
\end{equation*}
Then, for any
$ g_1, \dots, g_l \in \mathcal N^{m,0}$ 
and for each $ j = 1, \dots, l $,
\[
 f_j ( \vec y, \vec x ;  g_1 ( \vec x ; ), \dots,
                         g_l ( \vec x ; )
     )
 \in \mathcal N^{k+m,0}. 
\]
\end{thm}

Its analogue on SRN, 
which is called multiple predicative recursion on notation,
has been proved by Bellantoni:

\begin{thm}[Bellantoni \cite{bel92}] \label{thm1}

$ \mathcal B_{normal} $ 
is closed under simultaneous safe recursion on notation.
\end{thm}
\textbf{Proof of Theorem \ref{thm2}.}
In usual, simultaneous recursion is reduced to single recursion using 
a pairing function and unpairing functions.
Bellantoni also uses a pairing function
such that for 
$ a = ( a_1 a_2 a_3 )_2 $ and $ b = ( b_0 b_1 b_2 b_3 )_2 $,  
$ \langle a , b \rangle = ( a_3 b_3 a_2 b_2 a_1 b_1 0 b_0 )_2 $.
Let us recall its definition.
The pairing function $ \pi ( y ; a , b )$ is defined, by safe recursion on notation, by 
\begin{eqnarray*}
  \pi ( 0 ; a , b ) & = & 0 , \\
  \pi ( y i ; a , b ) & = & 
  C ( ; m ( y ; a ) ,
        C ( ; m ( y ; b ) , \pi ( y ; a , b ) 00  , \pi ( y ; a , b ) 01
    ) , 
\\ && \qquad \qquad \qquad \qquad 
        C ( ; m ( y ; b ) , \pi ( y ; a , b ) 10  , \pi ( y ; a , b ) 11 )
    )
\end{eqnarray*}
where
$ m ( y ; b ) $ is the $ | y | $th predecessor of $b$ 
($b$ \emph{minus} $y$ in unary notation) which is defined by
$ m ( 0 ; b ) = b $ and 
$ m ( y i ; b ) = P ( ; m ( y ; b ) ) $.
Then, for 
$ a = ( a_{ n-1 } \cdots a_0 )_2 $ and
$ b = ( b_{ m-1 } \cdots b_0 )_2 $,
\[
 \pi ( y , a , b ) =
 ( a_0 b_0 \cdots a_{ |y| - 1 } b_{ |y| - 1 } )_2 .
\]
Hence, if
$ f ( \vec x ; ) , 
  g ( \vec x ; ) \in \mathcal B_{normal}$, then 
\[
  \pi ( q ( \vec x ; ) ; 
        f ( \vec x ; ) , g ( \vec x ; )
      ) 
  = \langle f ( \vec x ; ) , 
            g ( \vec x ; ) 
    \rangle 
\]
where 
$ q \in \mathcal B_{normal} $
such that
$ 
 | f ( \vec x ; ) | , | g ( \vec x ; ) | 
 \leq | q ( \vec x ; ) | 
$.
In other words, we need to substitute a large enough value into the
position of a normal variable so that $\pi$ works as a pairing function.
As will be shown in the last section,
if $ f ( \vec x ; \vec a ) \in \mathcal N $, then
\[
 | f ( \vec x ; \vec a ) |
 \leq 
 2^{ p ( | \vec x | )
   }
 + \max | \vec a |
\]
for some polynomial $ p ( \vec x ) $.
If $ \max ( f ( \vec x ; ) , 
            g ( \vec x ; ) 
          ) \in 
     \mathcal N_{normal} \setminus \mathcal B_{normal} $,
i.e.,
$ \max ( f ( \vec x ; ) , 
         g ( \vec x ; ) 
       ) 
  \approx  
  2^{
      2^{ p ( | \vec x | ) 
        }
    }
$,
we have to substitute 
$
   2^{
      2^{ p ( | \vec x | ) 
        }
    }
$
into a normal position.
However, such composition is not allowed for us.  

Thus we define a ``high-speed'' pairing function 
$ \Pi_p ( \vec x ; a , b , c ) $ and the corresponding unpairing functions
$ \Pi_p^1 ( \vec x ; a , c ) $ and $ \Pi_p^2 ( \vec x ; a , c ) $
for \emph{each} polynomial $ p ( \vec x ) $.
$ \Pi_p $ and $ \Pi_p^j $ work as
\[
 \Pi_p ( \vec x ; a , b , c ) =
 c \oplus ( a_0 b_0 \cdots a_{ 2^{ p ( | \vec x | ) 
                                 } - 1
                             } b_{ 2^{ p ( | \vec x | ) 
                                     } - 1
                                 }
          )_2
\]
and
\begin{equation*}
 \Pi_p^j ( \vec x ; a , c ) =
 \begin{cases}
 c \oplus ( a_1 a_3 \cdots a_{ 2^{ p ( | \vec x | ) + 1
                                 } - 1 
                             }
          )_2
 & \text{ if $ j = 1 $, } \\
 c \oplus ( a_0 a_2 \cdots a_{ 2^{ p ( | \vec x | ) + 1
                                 } - 2
                             }
          )_2
 & \text{ if $ j = 2 $. }
 \end{cases}
\end{equation*}
Therefore, if
both $ f_1 ( \vec x ; ) $ and $ f_2 ( \vec x ; ) $ 
belong to $ \mathcal N_{normal}$ and
$ p ( \vec x ) $ is a polynomial such that
$
 | f_1 ( \vec x ; ) | , | f_2 ( \vec x ; ) | 
 \leq 
 2^{ p ( | \vec x | ) }
$, then
\[
 \Pi_p ( \vec x ; f_1 ( \vec x ; ), f_2 ( \vec x ; ) , 0
       ) 
  = \langle f_1 ( \vec x ; ), f_2 ( \vec x ; ) 
    \rangle
\]
and 
\[
 \Pi_p^j ( \vec x ;
           \Pi_p ( \vec x ; f_1 ( \vec x ; ) , 
                            f_2 ( \vec x ; ) , d
                 ) ,
           0
         )
 =
 f_j ( \vec x ; )
\]
for each $ j = 1 , 2 $ and an arbitrary $d$.
Simultaneously, we define
\begin{itemize}
\def\labelitemi{--}
\item
$ M_p ( \vec x ; a ) $,
the \emph{most} significant part of $a$,
which denotes the $ 2^{ p ( | \vec x | ) } $th predecessor of $a$,
\item
$ R_p ( \vec x ; a , c ) $
(the \emph{reverse} function),
which is 
$c$ concatenated with the right 
$ 2^{ p ( | \vec x | ) } $ bits of $a$ in reverse order, 
and 
\item $ L_p ( \vec x ; a ) $,
the \emph{least} significant part of $a$,
which denotes
the right $ 2^{ p ( | \vec x | ) } $ bits of $a$.
\end{itemize}
They are constructed step by step for polynomials
$ p_0 ( x ) = x, \ p_1 ( x , y , z ) = x \cdot y + z, \
  p_2 ( x , y , z , u , v , w ) 
  = x \cdot y \cdot z + u \cdot v + w , \dots $
along the construction in Example \ref{exam}.\ref{exam_0}--\ref{exam}.\ref{exam_2}. 
\begin{description}
\item[Step 1.]
We define $ \Pi_{ p_0 } ( y ; a , b , c ) $ and
$ \Pi_{ p_0 }^j ( y ; a , c )$.

$ M_{ p_0 } : = M, \ R_{ p_0 } : = R $ and $L_{ p_0 } : = L$
are defined by
\begin{equation*}
 \begin{cases}
  M ( 0 ; a ) = P ( ; a ),
 & \text{} \\
  M ( y i  ; a ) = M ( y ; M ( y ; a ) ) ,
 & \text{}
 \end{cases}
 \begin{cases}
  R ( 0 ; a , c ) = C ( ; a , c 0 , c 1 ), 
 & \text{} \\ 
  R ( y i ; a , c ) = 
  R ( y ; M ( y ; a ) , R ( y ; a , c ) ) ,
 & \text{}
 \end{cases}
\end{equation*}
and $ L ( y ; a ) = R ( y ; R ( y ; a , 0 ) , 0 ) $.

Then $ \Pi_{ p_0 } : = \Pi $ is defined by
\begin{eqnarray*}
   \Pi ( 0 ; a , b , c ) & = &
     C ( ; a , C ( ; b , c00 , c01 ) , 
               C ( ; b , c10 , c11 )
       ), \\
   \Pi ( y i ; a , b ,c ) & = & 
     \Pi ( y ; M ( y ; a ) , M ( y ; b ) , 
               \Pi ( y ; L ( y; a ), L ( y; b )
         )
       ) .
\end{eqnarray*}

And $ \Pi_{p_0}^j : = \Pi^j \ (j=1, 2) $
are defined by
\begin{eqnarray*}
   \Pi^1 ( 0 ; a , c ) & = & C ( ; P ( ; a ) , c0 , c1 ), \\
   \Pi^2 ( 0 ; a , c ) & = & C ( ; a , c0 , c1 ) \\ 
   \Pi^j ( y i ; a , c ) & = & 
     \Pi^j ( y ; M^2 ( y ; a ) , 
                 \Pi^j ( y ; L^2 ( y ; a ) , c 
                       )
           ) 
 \quad (i=0, 1) 
\end{eqnarray*}
where 
$ M^2 ( y ; a ) $ is 
the $ 2^{ |y| + 1 } $th predecessor of $ a $
defined by
$ M^2 ( y ; a ) = M ( y ; M ( y ; a ) ) $, and
$ L^2 ( y ; a )$ is the right $ 2^{ |y| + 1 } $ bits of $a$ 
defined to be
$ R^2 ( y ; R^2 ( y ; a , 0 ) , 0 ) $
for $ R^2 $ 
which is defined by
$ R^2 ( y ; a , c ) = 
  R ( y ; M ( y , a ) , R ( y ; a , c ) ) $.
\item[Step 2.]
We define
$ \Pi_{ p_1 } ( x , y , z ; a , b , c ) $ and 
$ \Pi_{ p_1 }^j ( x , y , z ; a , c ) $.

$ M_{ p_1 } : = M $ and  
$ R_{ p_1 } : = R $ are defined by
\begin{eqnarray*}
   M ( 0, 0, 0; a )
   &=&
   P (; a), \\
   M (x, y, z i; a)   
   & = & 
   M ( x, y, z ; M ( x, y, z ; a )), \\
   M ( x, y i, 0 ; a ) 
   & = & 
   M ( x , y, x; a ), \\
   M (xi, 0, 0; a) &=& M (x, 0, 0; a)
\end{eqnarray*}
and
\begin{eqnarray*}
   R (0, 0, 0; a, c)
   &=&
   C (; a, c0, c1), \\
   R ( x, y, z i ; a , c )
   & = & 
   R ( x, y, z ; M ( x, y, z ; a ) , 
                         R ( x, y, z ; a , c )
       ) , \\
   R ( x, y i , 0 ; a , c )
   & = & 
   R ( x , y, x ; a , c ), \\
   R (xi, 0, 0; a, c) &=& R (x, 0, 0; a, c).
\end{eqnarray*}
$ L_{ p_1 } $ is defined as $ L_{ p_0 } $.

Then $ \Pi_{ p_1 } : = \Pi $ is defined by
\begin{eqnarray*}
   \Pi (0, 0, 0; a , b , c ) 
   & = &
   \Pi_{p_0} (0; a , b , c ), \\
   \Pi ( x, y, z i ; a , b , c )
   & = & 
   \Pi ( x, y, z ; M ( x, y, z ; a ) ,
                         M ( x, y, z ; b ) , \\ 
   && \qquad \qquad 
                         \Pi ( x, y, z ; L ( x, y, z ; a ) ,
                                               L ( x, y, z ; b ) ,
                                               c 
                             )
       ) , \\
   \Pi ( x, y i , 0 ; a , b , c )
   & = &
   \Pi ( x , y, x; a , b , c ), \\
   \Pi (xi, 0, 0; a, b, c) &=& \Pi (x, 0, 0; a, b, c).
\end{eqnarray*}
Analogously, the functions $ \Pi_{ p_1 }^j $ are defined from 
$ M_{ p_1 }^2 $ and $ R_{ p_1 }^2 $. 
\end{description}

Given a polynomial $ p ( \vec x ) = p ( x_1 , \dots , x_k ) $, 
assume $ \Pi_{ p' }$ has been already constructed for a suitable 
polynomial $ p' ( x_1 , \dots , x_n )$
such that
\[
 p ( \vec x ) =
 p ' ( x_{ i_1 } , \dots , x_{ i_n } )
\]
for some 
$ i_1 , \dots , i_n \in \{ 1 , \dots , k \}$.
\begin{description}
\item[Final step.]
We apply a safe composition rule to 
$ \Pi_{ p' } ( x_1 , \dots , x_n ; a , b , c ) $ 
to get $ \Pi_p ( \vec x ; a , b , c ) $.
\end{description}

Now let us prove Theorem \ref{thm2}.
For simplicity,
consider the case $ l = 2 $ in the assertion.
Let 
$ p_g ( \vec x ), \ p_f ( \vec y, \vec x ) $
be polynomials such that for every $ j=1, 2 $,
\begin{eqnarray*}
 | g_j ( \vec x ; ) | 
 & \leq &
 2^{ p_g ( | \vec x | )
   }, \\
 | f_j ( \vec y, \vec x ; a_1, a_2 ) | 
 & \leq &
 2^{ p_f ( |\vec y|, | \vec x | )
   } + \max ( |a_1| , |a_2| ) .
\end{eqnarray*}
The canonical choice of such the polynomials $ p_g, \ p_f $ will be shown
in the proof of Lemma \ref{lem1}.
Put
$ p ( \vec y , \vec x ) := 
  p_f ( \vec y, \vec x ) + p_g ( \vec x )  
$ and
$ q ( \vec x ) := p ( \vec 0, \vec x ) $.
Then we define 
$ \hat f ( \vec y , \vec x ; a_1, a_2 ) $,
which is intended to be 
$ \langle f_1 ( \vec y , \vec x ; a_1 , a_2 ) , 
          f_2 ( \vec y , \vec x ; a_1, a_2 )
  \rangle
$,
by single SNRN equations such that
\[ \hat f ( \vec 0, \vec x ; a_1, a_2 ) = 
  \Pi_q ( \vec x ; 
                h_1 ( \vec x ; a_1, a_2 ) ,
                h_2 ( \vec x ; a_1, a_2 ) , 0
              ),
\]
and, in the case $ \max \vec y \neq 0$,
\begin{eqnarray*}
  \hat f ( \vec y, \vec x; \vec a) =
  \hat f ( \vec v_1, \vec x;
           \Pi^1_p (\vec v_2, \vec x; 
                        \hat f ( \vec v_2, \vec x; \vec a
                               ), 0
                   ), 
           \Pi^2_p (\vec v_2, \vec x; 
                        \hat f ( \vec v_2, \vec x; \vec a
                               ), 0
                   ) 
         ).
\end{eqnarray*}
By the definition of $ \Pi_p $ and $ \Pi^j_p $,
it can be shown that
\[
  | \hat f ( \vec y, \vec x; g_1 ( \vec x ), g_2 ( \vec x ; ) 
           ) 
  |
 \leq
 2 \cdot ( 2^{ p_f (| \vec y|, | \vec x|)  
             } + 2^{ p_g ( | \vec x | )
                   }
         )
 \leq
 2 \cdot
 2^{ p ( | \vec y|, | \vec x | ) 
   },
\] 
and, therefore,
\[
 f_j ( \vec y , \vec x ; g_1 ( \vec x ; ) ,
                         g_2 ( \vec x ; ) 
     ) 
 =
 \Pi_p^j ( \vec y , \vec x ;
           \hat f ( \vec y , \vec x ; g_1 ( \vec x ; ) ,
                                      g_2 ( \vec x ; )
                        ) ,
                 0
               )
 \in \mathcal N^{k+m,0} 
\]
for each $ j = 1 , 2 $.
\hfill $ \square $

\begin{col} \label{col} 

Suppose that $ f_1, \dots, f_l \in \mathcal N^{1+m,0} $
are defined from
$ g_1, \dots, g_l \in \mathcal N^{m,0} $ and
$ h_1, \dots, h_l \in \mathcal N^{m,l} $
simultaneously by safe recursion on notation such that
for each $ j = 1, \dots, l $,
\begin{equation*}
 \begin{cases}
  f_j ( 0 , \vec x ; ) = g_j ( \vec x ; ),
 & \text{} \\
  f_j ( y i , \vec x ; ) =
  h_j ( \vec x ; f_1 ( y , \vec x ; ), \dots , 
                 f_l ( y , \vec x ; )
      ) .
 & \text{} 
 \end{cases}
\end{equation*}
Then, for any polynomial $ p ( \vec x )$ and
for each $ j = 1, \dots, l $,
\[
 f_j ( 2^{ 2^{ p (|\vec x|) } } - 1 , \vec x ;  
     )
 \in \mathcal N^{m,0} .
\]
\end{col}
We notice that each $ h_j $ is independent of $ i = 0 , 1 $ and $y$. 

\begin{proof}
As in the previous proof,
consider the case $ l = 2 $.
Following the construction of $ M_p, R_p$ or $\Pi_p $,
we first define 
$ F_1 ( y , \vec x ; a_1, a_2 ) $ and $ F_2 ( y , \vec x ; a_1, a_2 ) $
such that 
\begin{equation}
 F_j ( y , \vec x ; g_1 ( \vec x ; ) , g_2 ( \vec x ; )
     )
 = f_j ( 2^{ 2^{ |y| }
           } - 1 , 
         \vec x ; 
       )
 \label{property}
\end{equation}
for each $ j = 1 , 2 $.
Using the simultaneous SNRN scheme in Theorem \ref{thm2},   
they are defined by
\begin{equation*}
 \begin{cases}
   F_j ( 0 , \vec x ; a_1, a_2 )
   = h_j ( \vec x ; a_1, a_2 ),
 & \text{} \\
   F_j ( y i , \vec x ; a_1, a_2 )
   = 
   F_j ( y , \vec x ; F_1 ( y , \vec x ; a_1, a_2 ) ,
                      F_2 ( y , \vec x ; a_1, a_2 ) 
       ).
 & \text{}
 \end{cases}
\end{equation*}
\begin{Claim}
For any $z$ and for each $ j = 1 , 2$,
\[
 F_j ( y , \vec x ; f_1 ( z , \vec x ; ) , 
                    f_2 ( z , \vec x ; ) 
     )
 = 
 f_j ( A_1 ( y ; z
           ) , \vec x ; 
     )
\]
where $ A_1 ( y ; a )$ is the $ 2^{ |y| } $th successor of $a$
with respect to $ S_1 $ 
(the \emph{addition} in unary notation) 
which is defined by 
$ A_1 ( 0 ; a ) = a 1 $ and 
$ A_1 ( y i ; a ) = A_1 ( y ; A_1 ( y ; a ))$.
\end{Claim}

In the claim, putting $ z = 0 $,
the desired property (\ref{property}) is enjoyed.
The claim is shown by simultaneous induction on $y$.  

In the case $ y = 0 $,
\begin{eqnarray*}
 F_j ( 0 , \vec x ; f_1 ( z , \vec x ; ) , 
                    f_2 ( z , \vec x ; )
     )
 & = & 
 h_j ( \vec x ; f_1 ( z , \vec x ; ) , 
                f_2 ( z , \vec x ; )
     )
 =
 f_j ( z 1 , \vec x ; ) \\
 & = & 
 f_j ( A_1 ( 0 ; z ) , \vec x ; ) .
\end{eqnarray*}
And in the case $ yi > 0 $,
\begin{eqnarray*}
 & &
 F_j ( y i , \vec x ; f_1 ( z , \vec x ; ) , 
                      f_2 ( z , \vec x ; )
     ) \\
 &=&
 F_j ( y , \vec x ; F_1 ( y , \vec x ; f_1 ( z , \vec x ; ) , 
                                       f_2 ( z , \vec x ; )
                        ) ,
                    F_2 ( y , \vec x ; f_1 ( z , \vec x ; ) , 
                                       f_2 ( z , \vec x ; )
                        )
     ) 
 \\
 &=&
 F_j ( y , \vec x ; f_1 ( A_1 ( y ; z ) , \vec x ; ) , 
                    f_2 ( A_1 ( y ; z ) , \vec x ; ) 
     ) 
 \quad \text{by the induction hypothesis} \\
 &=&
 f_j ( A_1 ( y ; A_1 ( y ; z )
            ) , 
       \vec x ; 
     )
 \quad \text{again by I.H.} \\
 &=&
 f_j ( A_1 ( y i ; z ) , \vec x ; ) .
\end{eqnarray*}
This concludes the claim.

By (\ref{property}) and Theorem \ref{thm2}, 
\[
 f_j ( 2^{ 2^{ |y| }
         } - 1 , 
       \vec x ; 
     )
 =
 F_j ( y , \vec x ; g_1 ( \vec x ; ) , g_2 ( \vec x ; ) 
     )
 \in \mathcal N^{ 1+k , 0 } .
\]

Next, as get 
$ M_{ p_1 } , R_{ p_1 }$ or $\Pi_{ p_1 } $,
we can define the functions
$  f_j ( 2^{2^{|y| \cdot |z| + |w|}} -1, \vec x;) $.
We observe that their definitions still satisfy the condition in Theorem 
\ref{thm2}, and hence,
\[
 f_j ( 2^{2^{|y| \cdot |z| + |w|}} -1, \vec x;)
 \in \mathcal N^{3+k,0}.
\]
Finally, by a suitable application of safe composition, we obtain
\[
 f_j ( 2^{2^{p(| \vec x|)}} -1, \vec x;)
 \in \mathcal N^{k,0}.
\]
\end{proof}

\section{EXPTIME functions belong to $ \mathcal N $}

In this section,
we show, 
with the use of Corollary \ref{col},
that every exponential-time computable function is a
member of $ \mathcal N_{normal} $.

\begin{thm} \label{thm3}
 If $f ( \vec x )$ is computed by a deterministic Turing machine
 within a number of steps bounded by 
 $ 2^{ p ( | \vec x |)}$ for some polynomial $p$,
 then $f( \vec x ; )$ belongs to $\mathcal N_{normal}$.  
\end{thm}
\begin{proof}
We simulate computations of a Turing machine by functions in 
$ \mathcal N $.
Assume the following one-tape Turing machine model 
$
 M= ( Q, \Sigma , \Gamma , \delta )
$.
\begin{itemize}
 \item $ Q = \{ q_0 , q_1 , \dots , q_m \}$ is a finite set of states,
       where $ q_1 , q_0 $ are the initial 
       and the halting state, resp.
 \item $ \Sigma = \{ 0 , 1 , B \}$ is a set of symbols.
       Each value is written in its binary representation from
       right to left on the tape.
 \item $ \Gamma = \{ \mathrm{ left , halt , right }\}$ 
       is a set of directions to which the head moves next.
 \item $ \delta : ( Q \setminus \{ q_0 \} ) \times \Sigma  
           \rightarrow Q \times \Sigma \times \Gamma $
       is the transition map for $ M $.
 \item In the initial state, the head scans the left next cell 
       to the left most symbol of inputs.
       In each step, 
       according to $ \delta $,
       the head rewrites the symbol scanned there 
       and moves left or right. 
       And when halts, it scans the right next cell to the right most
       symbol of the output. 
\end{itemize}

Let us encode the states, symbols and directions as
$ \lceil q_i \rceil = i $,
$ 
 \lceil 0 \rceil = 1 0 = 2, \lceil 1 \rceil = 1 1 = 3, 
\lceil B \rceil = 00 ,
 \lceil \mathrm{left} \rceil = 1 0 = 2, \lceil \mathrm{halt} \rceil = 0 , 
 \lceil \mathrm{right} \rceil = 1   
$,
and identify their code-numbers with themselves. 
Then we define some functions in $ \mathcal B $ which encode information
 on $M$ in step $ |t| $ of the computation on inputs 
$ \vec x = ( x_1 , \dots , x_n )$:

\begin{center}
  \begin{tabular}{rcp{8.5cm}}
  $ stat( t , \vec x ;) $ & $=$ & the state of $M$. \\
  $ symb ( t , \vec x ; ) $ & $=$ & the symbol which the head is
   scanning. \\
  $ direc ( t , \vec x ; ) $ & $=$ & the direction to which the head
       moves in the next step. \\
  $ left ( t , \vec x ; ) $ & $=$ & symbols from the left next to
       the symbol which the head is scanning, to the symbol on the left side
       of which only blank symbol $B$'s occur. \\ 
  $ right ( t , \vec x ; ) $ & $=$ & the same as  
       $ left ( t , \vec x ; )$ 
       except the word ``left'' replaced by ``right''.      
  \end{tabular}
\end{center}
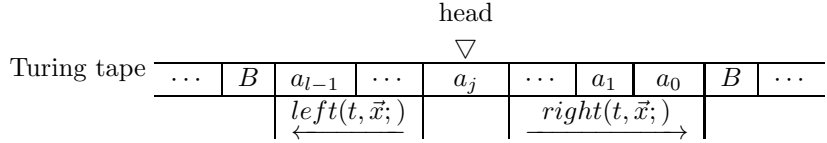
\begin{center}
  Turing tape
  \begin{tabular}{c|c|c|c|c|c|c|c|c|c}
  \multicolumn{4}{c}{} &
  \multicolumn{1}{c}{head} &
  \multicolumn{5}{c}{} \\
  \multicolumn{4}{c}{} &
  \multicolumn{1}{c}{$ \bigtriangledown $} &
  \multicolumn{5}{c}{} \\
  \hline
    $ \cdots $ & $B$ & $ a_{ l-1 } $ & $ \cdots $ & $ a_j $ &
    $ \cdots $ & $ a_1 $ & $ a_0 $ & $B$ & $ \cdots $ \\
  \hline
  \multicolumn{2}{c|}{} &
  \multicolumn{2}{|c|}{$ \underleftarrow{ left ( t , \vec x ; )
                                        }  $
                      } &
  \multicolumn{1}{|c|}{} &
  \multicolumn{3}{|c|}{$ \underrightarrow{ \mbox{ \ } right ( t , \vec x ; ) \mbox{ \ } } $} &
  \multicolumn{2}{|c}{}
  \end{tabular}
\end{center}

They are defined simultaneously by safe recursion on notation on $t$: \\
\underline{$ t = 0 $}
\begin{eqnarray*}   
    stat ( 0 , \vec x ; ) & = & \lceil q_1 \rceil = 1 \\ 
    symb ( 0 , \vec x ; ) & = & \lceil B \rceil = 0 \\
    direc ( 0 , \vec x ; ) & = & \lceil \text{right} \rceil = 1 \\
    left ( 0 , \vec x ; ) & = & 0 \\
    right ( 0 , \vec x ; ) & = & \oplus^n_n ( \vec x; )
\end{eqnarray*}
where $ \oplus^n_k \in \mathcal B^{ n, 0 } ( 0 \leq k \leq n ) $
is defined by induction on $k$ via an auxiliary function 
$ \oplus \in \mathcal B^{ 1, 1 } $.
The function $ \oplus ( x; a ) $, 
which denotes $ a \oplus \lceil B \rceil = a 00 $ followed by
$ \lceil x \rceil $ in reverse order,
is defined by 
$ \oplus ( 0; a ) = a 00 $ and 
$ \oplus ( xi ; a ) = \oplus ( x; a ) 1i $.
Then $ \oplus^n_k ( x_1, \dots, x_n;) ( k \leq n ) $ is defined by
\begin{equation*}
  \begin{cases}
    \oplus^n_0 ( \vec x; ) = O ( \vec x; ), 
  & \text{(the zero function)} \\
    \oplus^n_{ k+1 } ( \vec x ;) = 
    \oplus ( x_{ k+1 } ; \oplus^n_k ( \vec x ;) ).
  & \text{(safe composition)}   
  \end{cases}
\end{equation*}
By the definition, $ \oplus^n_k ( \vec x ; ) $ denotes the concatenation
 of the $k$ strings $ \lceil x_1 \rceil, \dots, \lceil x_k \rceil $
in reverse order with the string $00$ inserted.
Hence $ \oplus^n_n ( \vec x ;) $ denotes $ right ( 0, \vec x ; ) $:
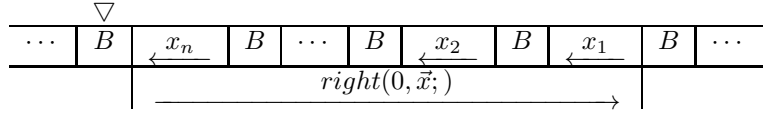
\begin{center}
  \begin{tabular}{c|c|c|c|c|c|c|c|c|c|c}
  \multicolumn{1}{c}{} &
  \multicolumn{1}{c}{$ \bigtriangledown $} &
  \multicolumn{9}{c}{} \\
  \hline
  $ \cdots $ & $B$ & 
  $ \underleftarrow{ \mbox{ \ } x_n \mbox{ \ } } $  & $B$ & $ \cdots $ & $B$
   & $ \underleftarrow{ \mbox{ \ } x_2 \mbox{ \ } } $ & $B$ & 
  $ \underleftarrow{ \mbox{ \ } x_1 \mbox{ \ } } $ & $B$ & $ \cdots $ \\
  \hline
  \multicolumn{2}{c|}{} &
  \multicolumn{7}{|c|}{$ \underrightarrow{ \mbox{ \qquad \qquad \qquad } 
                                           right ( 0 , \vec x ; )
                                           \mbox{ \qquad \qquad \qquad }
                                         }
                       $} &
  \multicolumn{2}{|c}{}
  \end{tabular}
\end{center}
\underline{$ t i > 0 $} \quad $ ( i = 0 , 1 ) $
\begin{eqnarray*}
  stat ( t i , \vec x ; ) & = & 
   \Delta_1 ( ; stat ( t , \vec x ; ) , symb ( t , \vec x ; ) ) \\
  symb ( t i , \vec x ; ) & = &
   \Delta_2 ( ; stat ( t , \vec x ; ) , symb ( t , \vec x ; ) ) \\
  direc ( t i , \vec x ; ) & = & 
   \Delta_3 ( ; stat ( t , \vec x ; ) , symb ( t , \vec x ; ) ) \\
  left ( t i , \vec x ; ) & = &
   \Delta_4 ( ; stat ( t , \vec x ; ) , symb ( t , \vec x ; ) , 
                direc ( t , \vec x ; ) , left ( t , \vec x ; )
             ) \\
  right ( t i , \vec x ; ) & = &
   \Delta_5 ( ; stat ( t , \vec x ; ) , symb ( t , \vec x ; ) , 
                direc ( t , \vec x ; ) , right ( t , \vec x ; )
             )
\end{eqnarray*}
where
$ \Delta_1 , \dots , \Delta_5 $
are defined according to the transition function $\delta$.
Since $\delta$ can be regarded as a finite function over natural numbers,  
we can easily convince ourselves that 
$ \Delta_1 , \dots , \Delta_5 $
are defined only on safe arguments using safe composition from 
initial functions.
\\

Suppose that $f ( \vec x ) $ is computed by $M$ within 
$ 2^{ p ( | \vec x |)}$-steps for some polynomial $p$.
Since 
$
 |
   2^{  2^{ p ( | \vec x | )
          }
     } -1
 |
 = 
 2^{ p ( | \vec x | )
   } 
$,
the values of 
$ stat ( t , \vec x ; ) , \dots , right ( t , \vec x ; ) $ on 
$
 t =
 2^{ 2^{  p ( | \vec x | )
       }
   } -1
$
are those at the time when the computation halts.
Moreover,
by the assumption on the position of the head of $M$ in its halting state,
$ left (  
         2^{ 2^{  p ( | \vec x | )
               }
           } -1 , \vec x ; 
        )$
encodes the value of $f ( \vec x )$.
The safe composition rule does \emph{not} allow to substitute 
$ 2^{2^{p(| \vec x|)}} -1 $
into a normal position. 
However, Corollary \ref{col} enables us to define
$ left (
         2^{ 
             2^{
                 p ( | \vec x | )
               }
           } -1 , \vec x ;  
       )
  \in \mathcal N_{normal}
$,
since, in the definitions of 
$ stat ( ti, \vec x ; ) , \dots , right ( ti, \vec x ; ) $,
$ \Delta_1, \dots, \Delta_5 $ depend neither on $ i = 0 , 1 $
nor on $t$.

Let 
$ Stat, \ Symb, \ Direc, \ Right, \ Left $
be defined respectively from
$stat$, $symb$, $direc$, $right$, $left$
as in Corollary \ref{col}, e.g.,
$ Left ( \vec x;) =
  left ( 2^{ 2^{  p ( | \vec x | )
               }
           } -1 , \vec x ; 
       )
$.
Namely, 
$ Stat ( \vec x;)$, $Symb ( \vec x;)$, $Direc ( \vec x;)$,
$Right ( \vec x;)$ and
$ Left ( \vec x ; ) $, respectively, encode 
\begin{itemize}
\def\labelitemi{--}
\item
the state of $M$ in step $ 2^{p (| \vec x|)}$, 
\item
the tape symbol scanned by $M$'s head in this step, 
\item
the direction $M$'s head moves in the next step,
\item
the tape inscription in this step read from the symbol right of the
 symbol scanned by the head to the symbol left of the first blank, and
\item
the tape inscription in this step read from the symbol left of the
 symbol scanned by the head to the symbol right of the first blank.
\end{itemize}
Let $ f ( \vec x ) = ( a_{ l-1 } \cdots a_0 )_2 $.
Then, in the halting state, the string 
$ a_{ l-1 } \cdots a_0 $ is written on the tape  as 
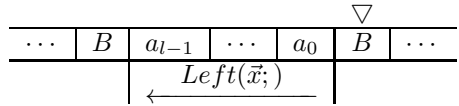
\begin{center}
\begin{tabular}{c|c|c|c|c|c|c}
\multicolumn{5}{c}{} &
\multicolumn{1}{c}{$ \bigtriangledown $} &
\multicolumn{1}{c}{} \\
\hline
$ \cdots $ &
$ B $ &
$ a_{ l-1 } $ &
$ \cdots $ &
$ a_0 $ &
$ B $ &
$ \cdots $ \\
\hline
\multicolumn{2}{c}{} &
\multicolumn{3}{|c|}{$ \underleftarrow{ \mbox{ \quad } 
                       Left ( \vec x ; ) \mbox{ \quad } } $
                    } &
\multicolumn{2}{c}{}
\end{tabular}
\end{center}
with the head of $M$ scanning the symbol $B$ next to $ a_0 $.
By the convention of our coding $ \lceil \cdot \rceil $,
$ Left ( \vec x ; ) = 
  ( 1 a_{ l-1 } \cdots 1 a_0 )_2 =
  \langle \underbrace{ 1 \cdots 1 }_{ l \text{-times} } , 
          ( a_0 \cdots a_{ l-1 } )_2
  \rangle  
$
whereas 
$ Right ( \vec x ; ) = 0 $.
Hence $ \Pi_Q^2 $ decodes $ f ( \vec x ) $ as
\[
 \Pi_Q^2 ( \vec x ; Left ( \vec x ; ) , d
         ) =
 d \oplus ( a_0 \cdots a_{ l-1 } )_2 
 \underbrace{0 \cdots 0}_{ 2^{ Q ( | \vec x | ) } - l
                         } .
\]
where $ Q ( \vec x ) $ is a length-bounding polynomial 
of $ Left ( \vec x ; )$ such that
\[
 | Left ( \vec x ; ) | = 2 l
 \leq
 2^{ Q ( | \vec x | ) + 1 } .
\]
Therefore, using the reverse function $ R_Q $ for the polynomial $Q$, 
we conclude
\[
 f ( \vec x ) =
 R_Q ( \vec x ; \Pi_Q^2 ( \vec x ; Left ( \vec x ; ), 0 ) , 0 )
 \in \mathcal N_{normal}.
\]
\end{proof}

If $f ( \vec x ) $ is computed by $M$ in 
$ p ( | \vec x | )$-steps for some polynomial $p$, then
we can define the above functions 
by simultaneous safe recursion on notation
using Bellantoni's pairing and unpairing functions in the previous section.
Hence a similar argument  will yield an alternative proof that
$ \mathcal F_{ \mathrm P } \subseteq \mathcal B_{normal}$.

\section{Functions of $\mathcal N$ are EXPTIME computable }
\label{lastsec}

The last section is devoted to show that every function in $\mathcal N$ is computed in 
exponential time on the lengths of inputs.
Using a standard technique, one can prove it.
For this  we need Lemma \ref{lem1} below.
We do not assume any particular machine model.
As mentioned in Remark \ref{rem2},
we prove it for a less restrictive scheme of SNRN.
Hence we define a subset $ \mathcal P ( \vec y, \vec x)$ of the set of
$\prec$-predecessors of $( \vec y, \vec x)$.

\begin{df}
\label{def4} \normalfont 
For $ k \geq 1 $ and $ m \geq 0 $,
let $ \vec y = ( y_1 , \dots , y_k )$ and
$ \vec x = ( x_1 , \dots , x_m )$.
Then the set $ \mathcal P ( \vec y, \vec x)$ is defined by
\[
 \mathcal P ( \vec y, \vec x) =
 \{ ( \vec v, \vec u) : ( \vec v, \vec u) \prec ( \vec y, \vec x) 
    \ \& \ \vec v \prec \vec y     
 \}.
\]
Furthermore, let
$ \Sigma^{k, m}_0 :=
  \Sigma^{k+m} \setminus 
  \{ \underbrace{ \z \cdots \z }_{k \text{ many}} \sigma_1, \cdots \sigma_m:
     \sigma_1, \dots, \sigma_m \in \Sigma
  \}
$.
Similarly to $ \Sigma^k_0$,
$ \Sigma^{k, m}_0 =
  \{ \tau ( \vec y, \vec x) : 
     \max \vec y \neq 0 \ \& \ x_1, \dots, x_m \geq 0 
  \}
$.

In this section, 
relaxing the definition of SNRN in Definition \ref{def3},
we mean by `SNRN' the scheme (\ref{snrn}) with $f$ depending on
$g$, $h_w$, $ \vec t_w$ and $ \vec s_w$ for every
$w \in \Sigma^{k, m}_0$, and also on some 
$ \prec^{k+m}$-functions $ \f_1$, $\f_2$ and $\f_3$ which induce 
$ J^{k+m}_{ \f_j ( \tau ( \vec y, \vec x))} ( \vec y, \vec x)
  \in \mathcal P ( \vec y, \vec x)
$
$(j=1, 2, 3)$ for all $( \vec y, \vec x)$ such that $ \max \vec y \neq 0$.
\end{df}
\textbf{Convention.}
From now on, 
for
$ \vec y = ( y_1 , \dots , y_k ) $ and
$ \vec x = ( x_1 , \dots , x_m ) $, we set
\[
 \sum ( d , \vec y , \vec x ) :=
 \sum_{ i=1 }^k ( \max ( | \vec y | , | \vec x | ) + 1 
                )^{ d-i } | y_i | +
 \sum_{ i=1 }^m ( \max ( | \vec y | , | \vec x | ) + 1 
                )^{ d-k-i } | x_i | .
\]

First we have a fundamental lemma on the predecessors.
Lemma \ref{lem_pred} is frequently used later. 
\begin{lem} \label{lem_pred}
If $ ( \vec v , \vec u ) \in \mathcal P ( \vec y , \vec x ) $ for
$ \vec y = ( y_1 , \dots , y_k ) $ and 
$ \vec x = ( x_1 , \dots , x_m ) $, then 
\[
 \sum ( d , \vec v , \vec u ) <
 \sum ( d , \vec y , \vec x )
\]
for all $ d \geq k+m $.
\end{lem}
\begin{proof}
Since all $ \vec v , \vec u $ are subterms of 
$ \vec y , \vec x $, first observe that
\[
 \max ( | \vec v | , | \vec u | ) \leq
 \max ( | \vec y | , | \vec x | ) .
\]
Let us recall the definition of $ \vec v \prec \vec y $
in Definition \ref{def2}.
Assume that $ v_1 = y_1 , \dots , v_{ n-1 } = y_{ n-1 } $,
$ v_n = P ( ; y_n ) $ 
for some $ n \leq k $. Then
$ | y_n | = | P ( ; y_n ) | + 1 = | v_n | + 1 $, and hence
\[
 c \cdot | v_n | + c = c ( | v_n | + 1 ) =
 c \cdot | y_n | 
\]
for any $c$.
Thus letting 
$ X = \max ( | \vec y | , | \vec x | ) + 1 $,
by the inequality
\[
 \sum_{ i = n+1 }^k ( \max ( | \vec v | , | \vec u | ) + 1
                    )^{ d - i } | v_i | +
 \sum_{ i = 1 }^m ( \max ( | \vec v | , | \vec u | ) + 1
                  )^{ d - k - i } | u_i | 
 < X^{ d - n } ,  
\]
we get 
\begin{eqnarray*}
 \sum ( d , \vec v , \vec u ) 
 & < & 
 \sum_{ i = 1 }^{ n - 1 } X^{ d - i } | v_i | + 
 X^{ d - n } \cdot | v_n | + X^{ d - n } \\
 & \leq & 
 \sum_{ i = 1 }^{ n - 1 } X^{ d - i } | y_i | + 
 X^{ d - n } \cdot | y_n |  \\
 & \leq &
 \sum ( d , \vec y , \vec x ) .
\end{eqnarray*}
\end{proof}

To prove the main theorem in this section,
we need Lemma \ref{lem1} which states that the length of every function in 
$ \mathcal N $ is bounded by some exponential in the lengths of the inputs.

\begin{lem} \label{lem1}
For any $ f \in \mathcal N^{k,l} $, there exists a constant $c$ such that
\[
  | f( \vec x ; \vec a ) | \leq
     2^{ c ( \sum_{ i = 1 }^k ( \max | \vec x | + 1 )^{ k - i }
             | x_i | + 1 
           )
       } 
             + \max | \vec a | .
\] 
\end{lem}

In the proofs of the lemma and Theorem \ref{thm4},
$ 2^{ p ( \vec x ) } $ will be written as 
$ \mathrm{ exp } ( p ( \vec x ) ) $.
\\ \\
\begin{proof}
We prove the lemma by induction over the construction of
$ f $.
The assertion is clear
if $f$ is any of initial functions.

For the induction step,
we deal only with the case SNRN.
The proof for safe composition is straightforward. 
For simplicity, 
suppose that
$ f \in \mathcal N^{k+m,l} $ is defined from
$g \in \mathcal N^{m,l}$,
$h_w$,
$t_{w, 1}, \dots, t_{w, l} 
 \in \mathcal N^{k+m,l+1}
$
$(w \in \Sigma^{k, m}_0)$,
and $\prec^{k+m}$-functions
$ \f_1$ and $ \f_2$ by
\renewcommand{\theequation}{\fnsymbol{equation}}
\setcounter{equation}{3} 
\begin{equation}
  \begin{cases}
  f ( \vec 0, \vec x; \vec a) = g ( \vec x; \vec a),
  & \text{} \\
  f ( \vec y, \vec x; \vec a) =
  h_{ \tau ( \vec y, \vec x)}
  ( \vec v_1, \vec u_1; \vec a,
    f ( \vec v_1, \vec u_1;
        \vec t_{ \tau ( \vec y, \vec x)}
        ( \vec v_2, \vec u_2; \vec a,
          f ( \vec v_2, \vec u_2; \vec a)
            )
      )
  )
  & \text{} \\
  ( \max \vec y \neq 0)
  & \text{}
  \end{cases}
\label{snrn2}
\end{equation}
where
$ ( \vec v_j, \vec u_j) :=
  J^{k+m}_{ \f_j ( \tau ( \vec y, \vec x))} ( \vec y, \vec x) 
  \in \mathcal P ( \vec y, \vec x)
$
for each $j=1, 2$.

Then, by I.H.,
there exist constants
$c_g$, $c_w$,
$c_{w, 1}, \dots, c_{w, l}$
respectively for
$ g$, $h_w$, $t_{w, 1}, \dots, t_{w, l}$
enjoying the condition.
Put
$ c := \max \{ c_g, c_w, c_{w, 1}, \dots, c_{w, l} :
               w \in \Sigma^{k, m}_0
            \} + 1
$.
Then by side induction on
$ \sum ( k+m, \vec y, \vec x ) $
we prove that
\[
  | f ( \vec y, \vec x; \vec a ) | \leq
  \mathrm{exp} ( c ( \sum ( k+m, \vec y, \vec x ) +1 )) + \max | \vec a | .
\]
In the base case $ \vec y = \vec 0 $,
by the main induction hypothesis for $g$,
\begin{eqnarray*}
           | f ( \vec 0, \vec x; \vec a ) | 
  & = &    | g ( \vec x; \vec a ) | \\
  & \leq & \mathrm{exp} ( c_g ( \sum ( k+m, \vec 0, \vec x ) +1 )) + \max | \vec a | \\
  & \leq & \mathrm{exp} ( c ( \sum ( k+m, \vec 0, \vec x ) +1 )) + \max | \vec a |.
\end{eqnarray*}
For the induction step, 
take an arbitrary 
$( \vec y, \vec x)$ such that $ \max \vec y \neq 0$.
By Lemma \ref{lem_pred},
\renewcommand{\theequation}{\arabic{equation}}
\setcounter{equation}{1}
\begin{equation} 
\label{pred}
 \sum ( k+m, \vec v_j , \vec u_j ) <
 \sum ( k+m, \vec y , \vec x) .
\end{equation}
Thus the side induction hypothesis yields 
\begin{equation}
 | f ( \vec v_j , \vec u_j ; \vec a ) |
 \leq
 \mathrm{ exp } 
 ( c ( \sum ( k+m, \vec v_j , \vec u_j ) + 1 
     )
 ) + \max | \vec a | .
\label{s.i.h.}
\end{equation}
Let
$h$, $t_i$ be 
$h_{ \tau ( \vec y, \vec x)}$, $t_{ \tau ( \vec y, \vec x), i}$, and
$c_h$, $c_i$ be
$c_{ \tau ( \vec y, \vec x)}$, $c_{ \tau ( \vec y, \vec x), i}$, resp.
Then, from M.I.H. for $t_i$, 
\begin{eqnarray*}
&      & | f ( \vec v_1, \vec u_1; 
               \vec t ( \vec v_2, \vec u_2; \vec a,
                        f ( \vec v_2, \vec u_2 )
                      )
             )
         | \\
& \leq & \mathrm{exp} ( c ( \sum ( k+m, \vec v_1, \vec u_1 ) +1)
                      ) + 
         \max \{ | t_i ( \vec v_2, \vec u_2; \vec a, 
                         f ( \vec v_2, \vec u_2; \vec a ) 
                       ) 
                 | : 1 \leq i \leq l
              \} \\
& \leq & \mathrm{exp} ( c ( \sum ( k+m, \vec v_1, \vec u_1 ) +1)
                      ) \\
&      & \quad +
         \max \{ \mathrm{exp} ( c_i ( \sum (k+m, \vec v_2, \vec u_2) +1
                                    ) 
                              )
                 + \max ( | \vec a |, | f ( \vec v_2, \vec u_2; \vec a ) |
                        )
                 : 1 \leq i \leq l
              \} \\
& \leq & \mathrm{exp} ( c ( \sum ( k+m, \vec v_1, \vec u_1 ) +1 )) \\
&      & \quad + 2 \cdot 
         \mathrm{exp} ( c ( \sum ( k+m, \vec v_2, \vec u_2 ) + 1 )) + 
         \max | \vec a | \quad \text{again by (\ref{s.i.h.})} \\
& \leq & 3 \cdot \mathrm{exp} ( c \sum ( k+m, \vec y, \vec x ))
         + \max | \vec a | \quad \text{by (\ref{pred})}.
\end{eqnarray*}
This together with M.I.H. for $h$ implies that
\begin{eqnarray*}
&    & | f ( \vec y, \vec x; \vec a ) | \\
& =  & | h ( \vec v_1, \vec u_1; \vec a, 
             f ( \vec v_1, \vec u_1; 
                 \vec t ( \vec v_2, \vec u_2; \vec a,
                          f ( \vec v_2, \vec u_2 )
                        )
               )
       | \\
&\leq& \mathrm{exp} ( c_h ( \sum ( k+m, \vec v_1, \vec u_1) +1 )) 
       + \max (| \vec a|, 
               | f ( \vec v_1, \vec u_1; 
                     \vec t ( \vec v_2, \vec u_2; \vec a,
                              f ( \vec v_2, \vec u_2 )
                            )
                   )
               | 
              ) \\
&\leq& \mathrm{exp} ( c_h ( \sum ( k+m, \vec v_1, \vec u_1) +1 )) +
       3 \cdot \mathrm{exp} ( c \sum ( k+m, \vec y, \vec x)) +
       \max | \vec a | \\
&\leq& 4 \cdot \mathrm{exp} ( c \sum ( k+m, \vec y, \vec x))
       + \max | \vec a | \\
&\leq& \mathrm{exp} ( c ( \sum ( k+m, \vec y, \vec x) + 1 )) 
       + \max | \vec a | \quad \text{by $ c \geq 2 $}.
\end{eqnarray*}
We notice that this proof is slightly extended to the case for an arbitrary
many times nested recursion.
This completes the proof of the lemma.
\end{proof} 

\begin{thm} \label{thm4}
If $ f \in \mathcal N^{ k, l } $, then
  $f( \vec x ; \vec a )$ is computable within a number of steps bounded by 
$ 2^{ c ( \sum_{ i = 1 }^k ( \max | \vec x | + 1 )^{ d - i }
          | x_i | + 1 
        )
    } 
  \cdot \max ( 2, | \vec a | ) 
$ for some constants 
$ c, d $.
\end{thm}

As a corollary of the theorem, our claim 
$ \mathcal N_{normal} \subseteq \mathcal F_{ \mathrm{ EXP } } $
follows.
\\ \\
\begin{proof}
We prove the theorem again by induction over the construction of
$ f $.
Let $ T_f ( \vec x ; \vec a ) $
be the least time needed to compute $ f ( \vec x; \vec a ) $.
If $f$ is any of initial functions,
it is clear since they are all linear-time computable.
The case that $f$ is defined by safe composition follows immediately
 from I.H. and Lemma \ref{lem1}.

For the case SNRN,
assume that $ f \in \mathcal N^{ k+m , l } $ is defined from
$g$, $h_w$, $t_{w, 1}, \dots, t_{w, l}$ $(w \in \Sigma^{k, m}_0)$
and $ \f_j$ $(j=1, 2)$
by the scheme (\ref{snrn2}) in the previous proof.
By Lemma \ref{lem1},
we have constants
$c_g'$, $c_w'$ and $c_{w, 1}', \dots, c_{w, l}'$
respectively for
$g$, $h_w$ and $t_{w, 1}, \dots, t_{w, l}$
enjoying the condition in the lemma.
As in the proof of the lemma, let
$c_0 := \max \{ c_g', c_w', c_{w, 1}', \dots, c_{w, l}' :
                w \in \Sigma^{k, m}_0
             \} + 1
$.

Furthermore, by I.H., 
there exist constants
$c_g$, $c_w$, $c_{w, 1}, \dots, c_{w, l}$ and
$d_g$, $d_w$, $d_{w, 1}, \dots, d_{w, l}$
respectively for
$g$, $h_w$, $t_{w, 1}, \dots, t_{w, l}$
enjoying the condition in the theorem.
Put
$c := 2 \max \{l+1, c_0, c_g, c_w, 
                    c_{w, 1}, \dots, 
                    c_{w, l} : w \in \Sigma^{k, m}_0
            \} 
$ and
$d := 2 \max \{ k+m, d_g, d_w, 
                    d_{w, 1}, \dots, 
                    d_{w, l} : w \in \Sigma^{k, m}_0
            \} 
$.
Now by side induction on
$\sum (d, \vec y, \vec x)$ 
we prove that
\[
 T_f ( \vec y, \vec x; \vec a ) \leq
  \mathrm{exp} ( c ( \sum ( d, \vec y, \vec x ) + 1 ))
  \cdot \max ( 2, | \vec a | ) .
\] 
Let $ \max_2 | \vec a | $ abbreviate $ \max ( 2, | \vec a | ) $.
The fact that $ \max_2 | \vec a | \geq 2 $ guarantees 
\begin{equation}
  \mathrm{exp} ( c ( \sum ( d, \vec y, \vec x ) +1)) + 
  \max | \vec a | \leq
  \mathrm{exp} ( c ( \sum ( d, \vec y, \vec x ) +1 ))
  \cdot \Max | \vec a | .
\label{max_2}
\end{equation}
In the base case, by M.I.H. for $g$,
\begin{eqnarray*}
 T_f ( \vec 0, \vec x; \vec a)
& =  & T_g ( \vec x; \vec a ) \\
&\leq& \mathrm{exp} ( c_g ( \sum ( k + d_g, \vec 0, \vec x ) +1 )
                    ) \cdot \Max | \vec a | \\
&\leq&
       \mathrm{exp} ( c ( \sum ( d, \vec 0, \vec x) +1 )
                    ) \cdot \Max | \vec a | .
\end{eqnarray*}
In the induction step, let us observe that for every 
$( \vec y, \vec x)$ such that $ \max \vec y \neq 0$,
\begin{eqnarray}
  T_f ( \vec y, \vec x; \vec a ) 
&=&
  T_h ( \vec v_1, \vec u_1; \vec a,
        f ( \vec v_1, \vec u_1; 
            \vec t ( \vec v_2, \vec u_2; \vec a,
                     f ( \vec v_2, \vec u_2; \vec a)
                   )
          )
      ) 
\nonumber \\
&& \quad +
  T_f ( \vec v_1, \vec u_1;
        \vec t ( \vec v_2, \vec u_2; \vec a,
                 f ( \vec v_2, \vec u_2; \vec a)
               )
      ) 
\nonumber \\
&& \qquad +
  \sum_{i=1}^l T_{t_i} ( \vec v_2, \vec u_2; \vec a,
                         f ( \vec v_2, \vec u_2; \vec a)
                       ) +
  T_f ( \vec v_2, \vec u_2; \vec a).
\label{Time_f}
\end{eqnarray}

Given $( \vec y, \vec x)$,
let $h$, $t_i$ be $h_{ \tau ( \vec y, \vec x)}$,
$t_{ \tau ( \vec y, \vec x), i}$,
$c_h$, $c_{i}$ be 
$c_{ \tau ( \vec y, \vec x)}$, 
$c_{ \tau ( \vec y, \vec x), i}$, and
$d_h$, $d_{i}$ be 
$d_{ \tau ( \vec y, \vec x)}$, 
$d_{ \tau ( \vec y, \vec x), i}$, resp.
First, the M.I.H. for $h$ yields that
\begin{eqnarray}
&&
  T_h ( \vec v_1, \vec u_1; \vec a,
        f ( \vec v_1, \vec u_1; 
            \vec t ( \vec v_2, \vec u_2; \vec a,
                     f ( \vec v_2, \vec u_2; \vec a)
                   )
          )
      ) 
\nonumber \\
&\leq& \mathrm{exp}
       ( c_h ( \sum ( d_h, \vec v_1, \vec u_1 ) +1 )
       ) \cdot \Max
       ( | \vec a|, 
         |f ( \vec v_1, \vec u_1; 
              \vec t ( \vec v_2, \vec u_2; \vec a,
                       f ( \vec v_2, \vec u_2; \vec a)
                     )
            )
         |
       )
\nonumber \\
&\leq& \mathrm{exp}
       ( c_h ( \sum ( d_h, \vec v_1, \vec u_1) +1 )
       )
       ( 3 \cdot
         \mathrm{exp}
         ( c_0 \sum ( k+m, \vec y, \vec x)
         ) + \max | \vec a |
       ) 
\nonumber \\
&\leq& 3 \cdot \mathrm{exp}
       ( c \sum ( d, \vec y, \vec x) 
       ) \cdot \Max | \vec a | \quad 
       \text{by (\ref{max_2}) and $ c \geq c_h + c_0 $}
\label{time_h}
\end{eqnarray}
where the second inequality has been shown in the proof of Lemma \ref{lem1}.
Next, by Lemma \ref{lem_pred} and S.I.H.,
\begin{eqnarray}
 T_f ( \vec v_j, \vec u_j; \vec a ) 
&\leq& 
       \mathrm{exp} ( c ( \sum ( d, \vec v_j, \vec u_j) +1 ) ) 
       \cdot \Max | \vec a | 
\nonumber \\
&\leq& 
      \mathrm{exp} ( c \sum ( d, \vec y, \vec x))
      \cdot \Max | \vec a | .
\label{time_f}
\end{eqnarray}
Hence this implies that
\begin{eqnarray}
 &    & T_f ( \vec v_1, \vec u_1; \vec t ( \vec v_2, \vec u_2; \vec a, 
                                      f ( \vec v_2, \vec u_2; \vec a )
                                    )
           ) 
\nonumber \\
&\leq& 
       \mathrm{exp} ( c ( \sum ( d, \vec v_1, \vec u_1) +1))
       \cdot \Max 
                       \{ |t_i ( \vec v_2, \vec u_2; \vec a, 
                                 f ( \vec v_2, \vec v_2; \vec a )
                               )
                          | : 1 \leq i \leq l
                       \}
\nonumber \\
&\leq& 
       \mathrm{exp} ( c (\sum ( d, \vec v_1, \vec u_1) +1))
          ( 2 \cdot \mathrm{exp} ( c_0 \sum ( k+m, \vec y, \vec x)
                                 ) + \max |\vec a| 
          ) 
\nonumber \\
&\leq& 2 \cdot
       \mathrm{exp} 
       ( c ( \sum ( d, \vec v_1, \vec u_1) + \sum ( k+m, \vec y, \vec x) +1
           )
       ) \cdot \Max | \vec a | 
       \quad \text{by (\ref{max_2})}
\nonumber \\
&\leq& 2 \cdot 
       \mathrm{exp}
       ( c \sum ( d, \vec y, \vec x)
       ) \cdot \Max | \vec a | 
       \quad \text{by $ d \geq 2 (k+m)$.}
\label{time_ff}
\end{eqnarray}
Similarly to the case $h$, by M.I.H. for $ t_i $, we have
\begin{eqnarray}
&    & \sum_{ i=1 }^l T_{ t_i } ( \vec v_2, \vec u_2; \vec a, 
                                   f ( \vec v_2, \vec u_2; \vec a )
                                 )
\nonumber \\ 
&\leq& \sum_{ i=1 }^l \mathrm{exp}
       ( c_i ( \sum ( d_i, \vec v_2, \vec u_2) +1 )
       )
       \cdot \Max 
         ( | \vec a |,
           | f ( \vec v_2, \vec u_2; \vec a )|
         )
\nonumber \\
&\leq& \sum_{ i=1 }^l \mathrm{exp}
       ( c_i ( \sum ( d_i, \vec v_2, \vec u_2) +1 )
       )
       ( \mathrm{exp}
         ( c_0 ( \sum ( k+m, \vec v_2, \vec u_2) + 1 )
         ) + \max | \vec a |
       )
\nonumber \\
&\leq& l \cdot \mathrm{exp}
       ( c \sum ( d, \vec y, \vec x)
       ) \cdot \Max | \vec a | 
       \quad \text{by $ c \geq c_i + c_0 $.}
\label{time_s}
\end{eqnarray}
Combining (\ref{Time_f}), (\ref{time_h}), (\ref{time_ff}), 
(\ref{time_s}) and (\ref{time_f}), we obtain
\begin{eqnarray*}
&    & T_f ( \vec y, \vec x; \vec a) \\ 
&\leq& (l+6)
       \mathrm{exp} ( c \sum ( d, \vec y, \vec x)) 
       \cdot \Max | \vec a | \\
&\leq& \mathrm{exp} (c \sum (d, \vec y, \vec x) +l+3)
       \cdot \Max | \vec a| \\
&\leq& \mathrm{exp} (c ( \sum (d, \vec y, \vec x) +1))
       \cdot \Max | \vec a|
       \quad \text{by $c \geq 2l + 2 \geq l+3$.}
\end{eqnarray*}
This completes the proof of the theorem.
We again notice that this proof works for the general form 
of SNRN.
\end{proof}

\section*{Acknowledgments}

The authors would like to thank the anonymous referees for careful reading.
They have pointed out some major errors and given us helpful comments and suggestions.
In particular, one of them has suggested a simpler scheme of SNRN. 
The present formulation (\ref{snrn}) in Definition \ref{def3} 
is due to him or her.
The second author would also like to thank Issei Shimizu for 
discussions with him and his encouragements at Kobe University. 
He pointed out that using simultaneous recursion, we can
arithmetize computations by Turing machines in $\mathcal N$ as in the
proof of Theorem \ref{thm3}.  


\end{document}